\documentclass[twocolumn,aps,prb,superscriptaddress,footinbib,floatfix]{revtex4-2}
\usepackage[utf8]{inputenc}

\usepackage{amsmath,amssymb,bm,ulem}
\usepackage{hyperref,xcolor,graphicx,siunitx}

\usepackage{tikzsymbols}
\usepackage{slashed}
\usepackage{ulem}

\begin{document}
\title{Bridging the small and large in twisted transition metal dicalcogenide homobilayers:  a tight binding model capturing orbital interference and topology across a wide range of twist angles.}
\author{Valentin Cr\'epel}
\affiliation{Center for Computational Quantum Physics, Flatiron Institute, New York, New York 10010, USA}
\author{Andrew Millis}
\affiliation{Center for Computational Quantum Physics, Flatiron Institute, New York, New York 10010, USA}
\affiliation{Department of Physics, Columbia University, New York, NY 10027, USA}

\begin{abstract}
Many of the important phases observed in twisted transition metal dichalcogenide homobilayers are driven by short-range interactions, which should be captured by a local tight binding description since no Wannier obstruction exists for these systems. 
Yet, published theoretical descriptions have been mutually inconsistent, with honeycomb lattice tight binding models adopted for some twist angles, triangular lattice models adopted for others, and with tight binding models forsaken in favor of band projected continuum models in many numerical simulations. 
Here, we derive and study a minimal model containing both honeycomb orbitals and a triangular site that represents the band physics across a wide range of twist angles. 
The model provides a natural basis to study the interplay of interaction and topology in these heterostructures. 
It elucidates from generic features of the bilayer the sequence of Chern numbers occurring as twist angle is varied, and the microscopic origin of the magic angle at which flat-band physics occurs. 
At integer filling, the model successfully captures the Chern ferromagnetic and van-Hove driven antiferromagnetic insulators experimentally observed for small and large angles, respectively, and allows a straightforward calculation of the magneto-electric properties of the system. 
\end{abstract}

\maketitle

\section{Introduction}

Interest in twisted transition metal dichalcogenide homobilayers (tTMDh) continues to increase as the number of correlated quantum phases experimentally observed in these heterostructures rapidly grows. Many of the observed phases, including Chern insulators~\cite{cai2023signatures,zeng2023integer}, fractional Chern insulators (FCI)~\cite{cai2023signatures,zeng2023integer,park2023observation,xu2023observation}, and van Hove-driven antiferromagnets~\cite{wang2020correlated,ghiotto2021quantum}, are  insulators characterized by a spontaneously broken symetry inherently driven by short-ranged interactions. For that reason, short-ranged descriptions of the bilayers using local tight binding models should help identify the important physics leading to the experimentally observed phase diagram, provide a basis for further calculations for example of transport and response functions, and enhance our understanding.  Yet, the tight binding descriptions of these materials have either been mutually inconsistent, being built on honeycomb/triangular lattices  for small/large twist angles respectively~\cite{devakul2021magic,zang2022dynamical}, or have missed the fine details of the experimentally measured phases, such as the robustness of the FCI with respect to twist angle variations~\cite{crepel2023anomalous,morales2023pressure}. In numerical simulations, tight binding approaches have been for the most part forsaken in favor of band projected continuum models, an approach that sometimes blurs the physical mechanism by which correlated phases emerge in these systems.

\begin{figure}
\includegraphics[width=\columnwidth]{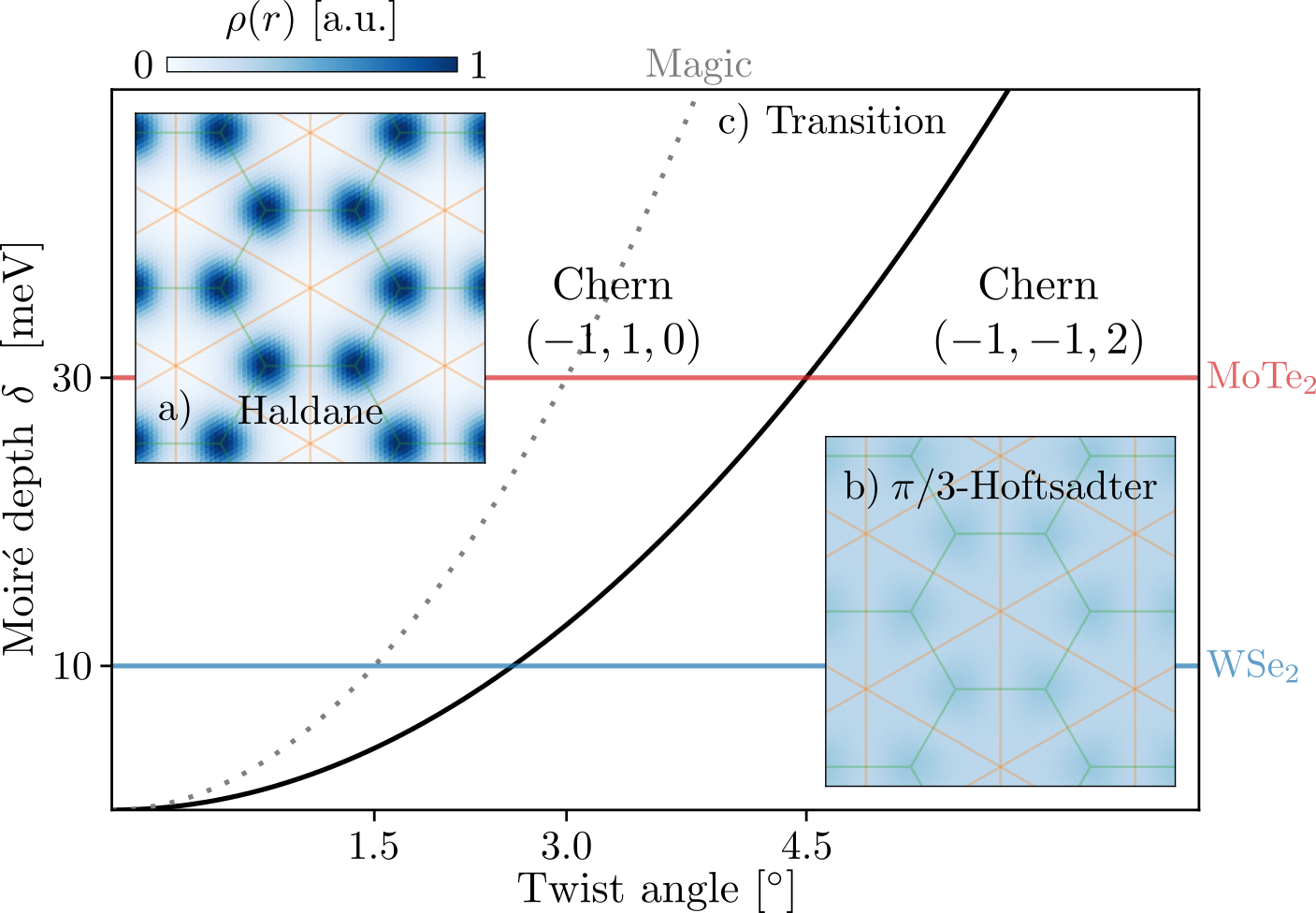}
\caption{Single-particle phase diagram of twisted transition metal dichalcogenide homobilayers in the plane spanned by the twist angle $\theta$, parameterizing the typical kinetic energy $E_{\rm kin} \propto \theta^2$ on the moir\'e scale, and the moir\'e depth $\delta$, measuring the potential energy difference between MX-XM (honeycomb) and MM (triangular) stacking sites. It shows a topological phase transition (black line) across which the spin-Chern numbers of the three topmost valence bands, counted from high to low energy, go from $(-1,1,0)$ to $(-1,-1,2)$ due to inversion between the second and third hole bands; and a ``magic'' line (dotted) where the topmost valence band becomes extremely flat. Blue and red horizontal lines indicate estimates of $\delta$ for WSe$_2$ and MoTe$_2$ respectively in this range of $\theta$~\cite{devakul2021magic,wang2023fractional}. These values are non-universal, \textit{e.g.} they depend on lattice-relaxation, and should only serve as prototypical references. 
a) Top left inset: charge density $\rho(r)$ of the topmost hole band in the $\delta \gg E_{\rm kin}$ regime, showing charge localization at the honeycomb sites. In this limit, the system realizes a time-reversal symmetric Haldane model. 
b) Lower right inset: same as (a) in the $\delta \ll E_{\rm kin}$ regime, showing equal occupancy of honeycomb and triangular sites. In this limit, the system realizes a time-reversal symmetric $\pi/3$-Hofstadter model.}
\label{fig_sketchresults}
\end{figure}

This paper uses very general symmetry considerations to derive a local tight binding model (\textit{i.e.} with short ranged hoppings) that encodes all of the known single particle physics of tTMDh. The model involves three orbitals per unit cell, two centered on the sites of a hexagonal lattice and one centered on the ``triangular" sites (hexagon center), resulting in three bands whose orbital composition and band topology (Chern number) depends on the twist angle and interlayer potentials. The necessity for a three orbital tight binding model comes from the following observations. On the one hand, the regions of the moir\'e pattern with lowest electrostatic energy have a local MX or XM stacking configuration~\cite{carr2018relaxation} (where a chalcogen atom X in one layer almost perfectly aligned with a metallic atom M of the opposite layer), which form a honeycomb lattice shown in green in Fig.~\ref{fig_sketchresults} and~\ref{fig_bilayer}a. On the other hand, interlayer tunneling at these points of charge accumulation must vanish due to the $C_3$-symmetry of the hole-carrying metallic orbitals~\cite{tong2017topological}. For the same reason, interlayer tunneling is found to be strongest in the MM stacking region (where metallic atoms of both layers overlay), which form a triangular lattice shown in orange in  Fig.~\ref{fig_sketchresults} and~\ref{fig_bilayer}a. The structure of the highest-lying valence bands of tTMDh involves an interplay between these two effects, which requires a three orbital model with one triangular and two hexagonal sites per moir\'e unit cell.

Another important question for tTMDh concerns the existence, generality and the value of ``magic'' angles at which the topmost valence band becomes extremely flat. Such angles were theoretically observed to appear at small angles in effective continuum descriptions of the bilayers~\cite{wu2019topological,devakul2021magic}, but general lattice-based arguments ensuring their existence have not been presented. This should be contrasted with twisted bilayer graphene, where symmetry arguments ensure that the Fermi velocity can be made vanishing by tuning a single parameter of the theory~\cite{sheffer2023symmetries}, demonstrating the generic existence of magic angles. Because monolayer TMDs may be regarded as gapped Dirac materials~\cite{liu2015electronic}, the arguments used for graphene~\cite{tarnopolsky2019origin,becker2022mathematics,crepel2023chiral2,estienne2023ideal} were recently extended to tTMDh. This approach allows to interpret the flat-band as a Landau level arising from pseudo-magnetic fields and has lead to values of the magic angle compatible with experimental measurements~\cite{crepel2023chiral,morales2024magic,crepel2024topologically}. Whether observed in continuum model or derived from the underlying Dirac physics of the TMDs, these magic angles have only been described in tight binding models through long-ranged hoppings~\cite{devakul2021magic,crepel2023anomalous}, which does not fit naturally with the presumed short-ranged origin of the experimentally observed phases. 

We use our three orbital model to present a general single-particle phase diagram for tTMDh, relying solely on the geometry of the bilayer, the spin-valley locking within each monolayer, and the scaling of kinetic energy with twist angle. Our theory, summarized schematically in Fig.~\ref{fig_sketchresults}, simultaneously ($i$) interpolates between the small and large twist angle regimes and naturally explains the sequences of Chern numbers observed in these two limits; ($ii$) justifies the existence of a magic angle as a result of orbital interference at the single-particle level; and ($iii$) captures the short-ranged correlated phases observed experimentally using local interactions. It therefore provides a natural local basis to study the interplay between interaction and topology, the magneto-electric couplings that take place in tTMDh, and describes fluctuation and correlation effects beyond the reach of continuum mean field theory or small system numerics.

Using this description of the bilayer, we identify other features of potential interest. For instance, the interference effects flattening the topmost valence band can also flatten the second highest valence band when it is almost fully filled. This suggests that similar correlated physics occurs at the filling of $x<1$ and $4-x$ holes per moir\'e unit cells, and could motivate new experimental studies at higher doping concentrations.

The rest of this paper is organized as follows. Section~\ref{sec:effectivemodel} presents a derivation of the effective model, describes its topology in relation to ab-initio continuum models, discusses the associated Wannier orbitals and gives representative band and interaction parameters. Section~\ref{sec_limitingcases} analyses the  structure of the model, showing how the sequence of band Chern numbers, the topological transition, and the existence of a magic angle at which the top band is nearly flat follow from qualitative considerations and are generic features of the model. Section~\ref{sec:correlatedphases} uses a Hartree-Fock analysis of the tight binding model to construct a range of ordered phases and deduce interesting magneto-electric effects. Section~\ref{sec:conclusion} gives a summary of our main results. Appendices give details of some of the calculations.

\section{Effective three-orbital model \label{sec:effectivemodel}} 

In this section, we use the symmetries of the tTMDh and the strong spin-valley locking inherited from the constituent monolayers to derive the generic form of a tight binding Hamiltonian for a tTMDh with twist angle $\theta$ and moir\'e lattice constant $a_m$. The model involves three orbitals; two at the hexagonal (XM/MX) and one at the triangular (MM) sites of the Moire uni cell. The derivation relies only on the assumption that the three orbitals identified above have an $s$-wave character (in other words, with the associated charge density transforming as the identity representation under the point symmetries associated with the orbital location). For continuum models, the strength of the hopping terms can be efficiently determined by Wannierization through maximal layer projection and fitting~\cite{devakul2021magic,qiu2023interaction}.

Keeping only terms of range $a_m$ in this model already accounts for the Chern numbers of the three topmost valence bands observed in large-scale \textit{ab-initio} calculations~\cite{wang2023fractional,jia2023moir} and continuum models developed for hole-doped tTMDh~\cite{wu2019topological,wang2023fractional,reddy2023fractional}. 
Keeping also terms of range $ 2 a_m$ allows to reproduce the energetic of the topmost two bands with high accuracy.

\subsection{Construction of the model}

\begin{figure}
\includegraphics[width=\columnwidth]{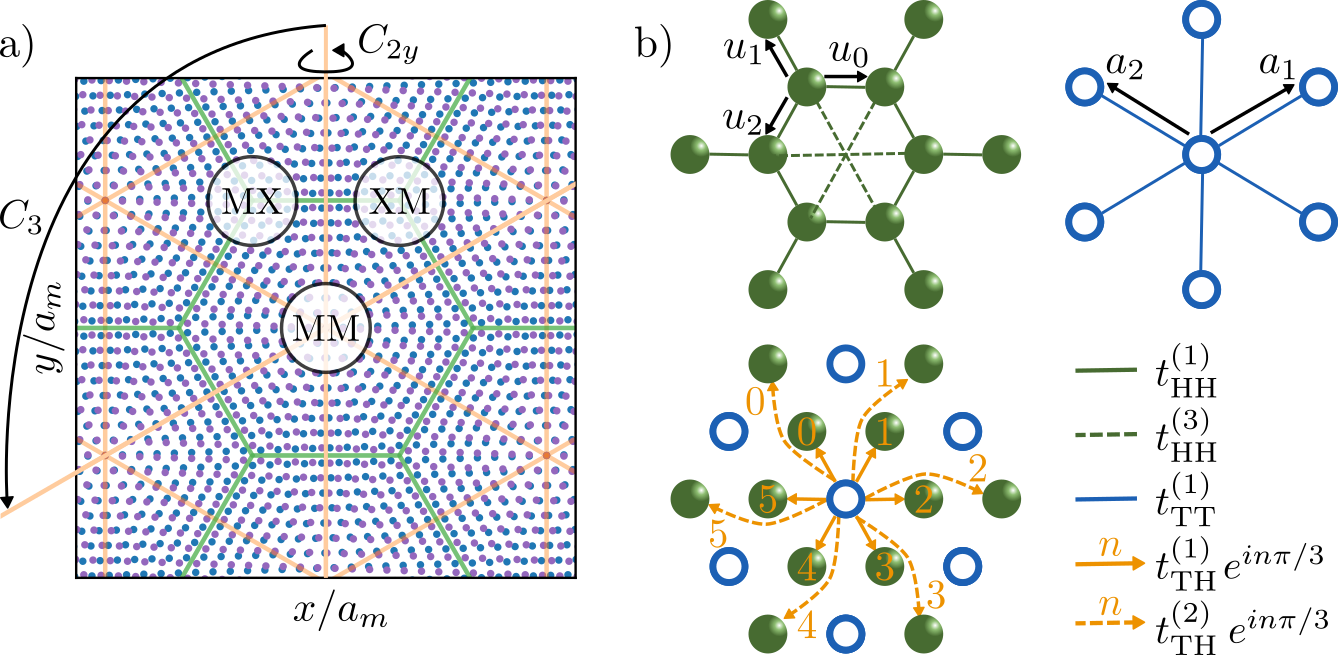}
\caption{a) Sketch of unit cell of a twisted transition metal dichalcogenides homobilayer, showing the high symmetry (MX/XM and MM) stacking regions where hole charge density is concentrated, and the action of the $C_3$ (rotation about an axis perpendicular to the plane) and $C_{2y}$ (rotation by $\pi$ about an axis in the plane) symmetries.
b) Representation of the five non-zero tunneling amplitudes of the three-orbital model in the $\tau=+$ valley. The phases of the hopping parameters are fixed by the momentum of spin-valley locked holes and the geometry of the bilayer. 
}
\label{fig_bilayer}
\end{figure}

We consider $p$-doped tTMDh, schematically depicted in Fig.~\ref{fig_bilayer}a, which shows the atomic positions, the moir\'e unit cell boundaries, the high-symmetry points where the hole charge density is concentrated and the action of the $C_3$ (rotation about an axis perpendicular to the plane) and $C_{2y}$ (rotation by $\pi$ about an axis in the plane) symmetries. The holes in each layer are spin-valley locked due to a strong Ising spin-orbit coupling~\cite{xiao2012coupled}, such that spin $\uparrow$/$\downarrow$ holes have their momentum located around the $\tau K$ corner of the monolayer's Brillouin zone, with $\tau = +$/$-$. The single-particle Hamiltonian of spin-$\downarrow$ in valley $K'\equiv -K$ is deduced from that of spin-$\uparrow$ in valley $K$ using time-reversal conjugation.

The argument outlined in the introduction suggests the construction of a local tight binding description of the tTMDh using three orbitals, two of them centered on the XM-MX (honeycomb) lattice positions enjoying the lowest electrostatic energy within the moir\'e unit cell, and one centered on the MM triangular lattice where inter-layer tunneling is the strongest. We observe that the calculated DFT or continuum model band structures exhibit three high-lying bands with zero net Chern number~\cite{wu2019topological,devakul2021magic,wang2023fractional,yu2023fractional}, so that there is no topological obstruction to a representation involving three exponentially localized Wannier functions. Our only assumption is that these three orbitals have an $s$-wave character, so that the phases of the hopping parameters in our model solely come from the dynamical phase acquired by spin-valley locked holes due to their non-zero center of mass momentum. These phases only apply to intra-layer tunnelings mediated by the in-plane kinetic terms, inter-layer tunneling on the other hand are predominantly local and remain real. The bilayer geometry fixes the layer-polarization of our three orbitals, and therefore fully determines which hopping processes acquire such phases. Indeed, holes are carried by the $d_{x^2-y^2} +i \tau d_{xy}$ orbital combination of the metal atom. As a result, exactly at the XM/MX points, the low energy holes are entirely polarized in the bottom/top layer, and so must our tight binding orbital be. The MM triangular site is related to itself by the $C_{2y}$ symmetry and must therefore have equal amplitude on each layer, as expected from an inter-layer bonding orbital at the position of maximal interlayer tunneling. 

The spin-valley locking and layer-polarization of the orbitals deduced from the geometry of the bilayer completely specify the phases in our three orbital model, which are shown in Fig.~\ref{fig_bilayer}b. For instance, nearest-neighbor tunneling between honeycomb sites is purely inter-layer and therefore real, while nearest-neighbor tunneling between triangular and honeycomb sites are mediated by intra-layer terms and possess phases determined to be multiples of $\pi/3$ by $C_3$ and $C_{2y}$ symmetry~\cite{rademaker2022spin} (which correspond to $e^{i \tau K \cdot d}$ for hole in valley $\tau$ hopping by a distance $d$). Similar considerations for next-nearest neighbors lead to the second-neighbor terms depicted in Fig.~\ref{fig_bilayer}b; farther distance hppings may be constructed similarly but are not shown here.

Let us give a more formal description of these terms. We focus on the spin-$\uparrow$ (valley $\tau=+$) component, and combine the three orbitals into a spinor $\Psi_k = [c_{{\rm MX},k}, c_{{\rm MM},k}, c_{{\rm XM},k}]^T$ with $\{ {\rm MX}, {\rm MM}, {XM} \}$ labeling the three orbitals shown in Fig.~\ref{fig_bilayer}a, $k \in {\rm mBz}$ a momentum index, mBz the moir\'e Brillouin zone, and $c_{o,k}$ the fermionic operator annihilating a particle with momentum $k$ on orbital $o$. In this basis, the three orbital tight binding model sketched in Fig.~\ref{fig_bilayer}b takes the form 
\begin{equation} \label{eq_fullTB}
\mathcal{H}_{\uparrow} (k) = \Psi_k^\dagger \left[ h_{\rm pot} + h_{\rm nn}(k) + h_{\rm lr}(k) \right] \Psi_k , 
\end{equation}
where $h_{\rm pot}$ contains all constant potential terms, while the tunneling terms are split into two groups: the dominant ones with range $a_m$ are gathered in $h_{\rm nn}$, all longer range ones are collected in $h_{\rm lr}$. They read
\begin{widetext}
\begin{equation} \label{eq_expandedfullTB}
h_{\rm pot} = \begin{bmatrix} E_z &0 & 0 \\ 0& -\delta&0 \\ 0&0&-E_z \end{bmatrix} , \;\;\;
h_{\rm nn} (k ) = \begin{bmatrix} 0 & t_{\rm TH}^{(1)} g_{k} & t_{\rm HH}^{(1)} f_{k}^* \\ t_{\rm TH}^{(1)} g_{k}^*& 0&-t_{\rm TH}^{(1)} g_{-k}^* \\ t_{\rm HH}^{(1)} f_{k}&- t_{\rm TH}^{(1)}g_{-k}&0 \end{bmatrix} ,  \;\;\; 
h_{\rm lr} (k ) = \begin{bmatrix} 0 & t_{\rm TH}^{(2)} g_{2k} & t_{\rm HH}^{(3)} f_{2k} \\ t_{\rm TH}^{(2)} g_{2k}^*& t_{\rm TT}^{(1)} h_k &-t_{\rm TH}^{(2)} g_{-2k}^* \\ t_{\rm HH}^{(3)} f_{2k}^* &- t_{\rm TH}^{(2)}g_{-2k}&0 \end{bmatrix}  ,
\end{equation}
with 
\begin{equation}
f_{k} =  \sum_{j=0}^2 e^{i (k \cdot u_j)} ,  \quad g_{k} = \sum_{j=0}^2 e^{2i\pi (1-j) / 3} e^{i (k \cdot u_j)} , \quad h_k = 2 \sum_{j=1}^3 \cos (k\cdot a_j) . 
\end{equation}
\end{widetext}
The definitions of the tunneling coefficients and of the vectors $u_{j=0,1,2}$ and $a_{j=1,2,3}$ are given graphically in Fig.~\ref{fig_bilayer}b. The term $\delta$ in $h_{\rm pot}$ measures the potential difference between the honeycomb and triangular sites, which corresponds to the moir\'e depth represented in Fig.~\ref{fig_sketchresults}. The second coefficient $E_z$ corresponds to the effect of an out-of-plane electric field, which introduces a difference of potential between the top and bottom layers but does not affect the MM orbital to linear order in $E_z$ since the latter has equal amplitude on both layers.

\subsection{Topological content compared to \textit{ab-initio}}

Because the studied tTMDh has $C_3$ rotational symmetry, the Chern number of its topmost valence bands can be determined modulo three by the $C_3$-eigenvalues of the Bloch states at the center $\gamma$ and corners $\kappa/\kappa'$ of the mBz~\cite{fang2012bulk}. In our tight binding model, each orbital gives one eigenvalue at these high-symmetry point as the product of two contributions (we discard the spin factor): the first comes from the angular momentum of the metallic $d_{x^2-y^2}+i\tau d_{xy}$ orbitals, and the second arises from the possible shift introduced by $C_3$ combined with the center of mass momentum $\tau K$ of spin-valley locked holes. For spin-$\uparrow$ holes, we find $(1,\omega,\omega)$ for the three eigenvalues at $\gamma$ and $(\omega, \omega, \omega^* )$ at $\kappa/\kappa'$, where $\omega = e^{2i\pi/3}$~\cite{luo2023symmetric}. Our tight binding model Eq.~\ref{eq_fullTB} can only faithfully describe the topology of the three topmost bands of the tTMDh if the three topmost spin-$\uparrow$ band of the latter possess the same triplet of eigenvalues at $\gamma$, $\kappa$ and $\kappa'$. The precise order in which they appear depends on the energetics of the model, \textit{i.e.} the sign and amplitudes of the parameters in Eq.~\ref{eq_expandedfullTB}.

Large-scale \textit{ab-initio} calculations of twisted MoTe$_2$ homobilayers have been able to compute these $C_3$ eigenvalues at certain commensurate twist angles, finding for spin-$\uparrow$ holes~\cite{wang2023fractional,private}
\begin{equation} \label{eq_c3eigvals}
\begin{array}{@{}c||ccc|c@{}}
{\rm band} & \gamma & \kappa & \kappa' & C \, ({\rm mod} \, 3) \\ 
\cline{1-5}\noalign{\vskip\doublerulesep\vskip-\arrayrulewidth}\cline{1-5}
1 & 1 & \omega & \omega & -1 \\
2 & 1 & \omega^* & \omega^* & 1 \\
3 & \omega & \omega & \omega & 0
\end{array} , \quad  
\begin{array}{@{}ccc|c@{}}
\gamma & \kappa & \kappa' & C \, ({\rm mod} \, 3) \\ 
\cline{1-4}\noalign{\vskip\doublerulesep\vskip-\arrayrulewidth}\cline{1-4}
1 & \omega & \omega & -1 \\
\omega & \omega^* & \omega^* & -1 \\
1 & \omega & \omega & 2 
\end{array} 
\end{equation}
respectively for small and large twist angles, where the three topmost valence bands are counted from high to low energies. We have also indicated the Chern number compatible with these eigenvalues~\cite{fang2012bulk}. The same sequence of Chern numbers is observed in continuum models for twisted WSe$_2$~\cite{devakul2021magic} and MoTe$_2$~\cite{wu2019topological,reddy2023fractional,jia2023moir}, with the exact value of the twist angle separating the two regimes of Eq.~\ref{eq_c3eigvals} differing from model to model. Both sets of eigenvalues in Eq.~\ref{eq_c3eigvals}, and hence the band topology obtained in \textit{ab-initio} calculations, are compatible with our tight binding model.

Turning to the energetics of our model, we can reproduce the Chern sequence of Eq.~\ref{eq_c3eigvals} using simple limits of our model with nearest neighbor tunneling only. To reproduce the topology observed for large twist angles, we can for instance set $t_{\rm TH}^{(1)} = t_{\rm HH}^{(1)} >0$ and all other parameters to zero. This reduces the problem to a $\sqrt{3}\times\sqrt{3}$-reduced triangular lattice in which each triangle is threaded by a $\pi/3$-flux. The resulting $\pi/3$-Hofstadter model has the desired sequence of Chern number $(-1,-1,2)$. For small twist angles, we additionally turn on a large $\delta \gg t_{\rm TH}^{(1)}$ and adiabatically eliminate the highly energetic triangular orbital. This provide a model on the honeycomb lattice with complex next-nearest hoppings of the Haldane type~\cite{haldane1988model}, leading to the Chern number sequence $(-1,1,0)$. We will see that these two regimes are, in fact, representative of the physics of tTMDh in Sec.~\ref{sec_limitingcases}, as quantitatively determined by fitting the tight binding parameters to the band structure of effective continuum models of tTMDh.

\subsection{Wannierization by layer polarization}

Our knowledge of the orbital symmetry and location allows us to connect the tight binding orbitals defined on general grounds above to a straightforward Wannierization of the three topmost valence of effective continuum models for tTMDh with free parameters fitted to large-scale \textit{ab-inito} band structures. Indeed, we know that holes are mostly carried by the metallic atoms in TMDs, such that the orbitals found at MX and XM (Eq.~\ref{eq_c3eigvals}) must show strong layer polarization. These orbitals being related by $C_{2y}$ symmetry, their layer polarization must furthermore be exactly opposite. Similarly, the triangular orbital at MM is related to itself by $C_{2y}$ and will be layer unpolarized. Diagonalizing the layer-polarization operator shall therefore provide the three Wannier orbitals identified from symmetry considerations above. 

We follow Refs.~\cite{devakul2021magic,qiu2023interaction}, and construct our three Wannier orbitals as follows: ($i$) at each momenta in the mBz, we diagonalize the layer polarization operator in the subspace spanned by the Bloch eigenvectors of the three first bands, and ($ii$) we Fourier transform the obtained layer-resolved states to find the Wannier orbitals in real-space. The gauge of the layer-resolved states is fixed from our knowledge of the orbital symmetry: we require the layer-resolved states to be real at the position of their designated center, in agreement with the $s$-wave orbital symmetry assumed above. 

\begin{figure*}
\includegraphics[width=0.95\textwidth]{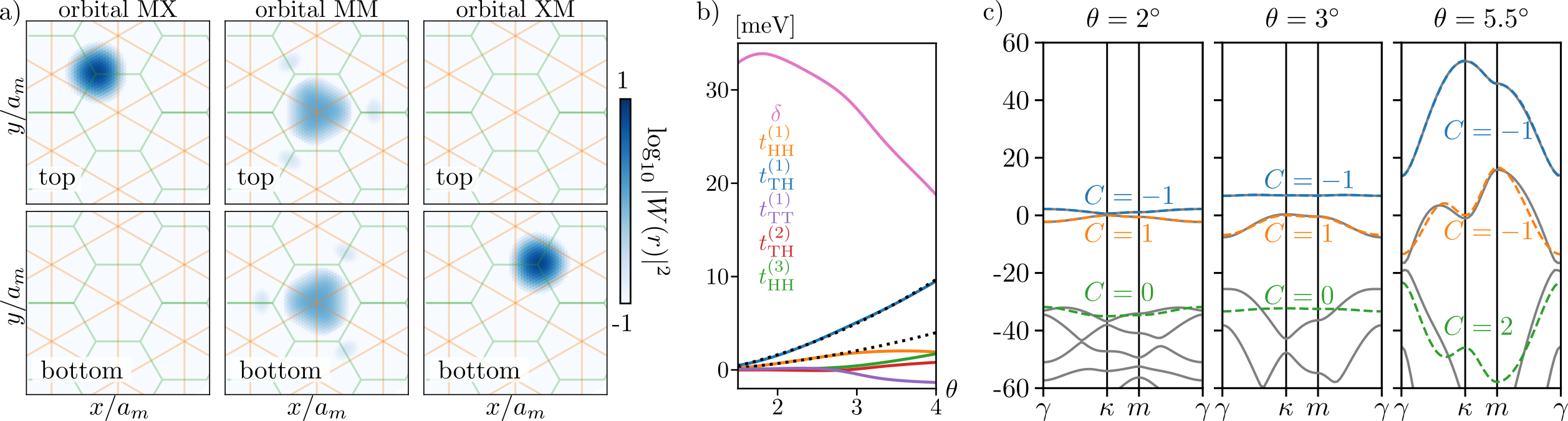}
\caption{
a) Logarithm of the square modulus of the three Wannier orbitals obtained by Fourier transforming the layer polarization eigenstates (columns) for the continuum model of Ref.~\cite{wang2023fractional} at $\theta=3^\circ$. The top/bottom row correspond to the amplitude in the top/bottom layer.
b) Fitted tight binding parameters of the three orbital model as a function of twist angle $\theta$ for the continuum model of Ref.~\cite{wang2023fractional}.
c) Band structure from the continuum model extrapolating the \textit{ab-initio} calculations of Ref.~\cite{wang2023fractional} to all angles (gray), compared to the results of the three-orbital tight binding model (dashed colors) including all tunneling of range $\leq 2a_m$. The Chern numbers of the topmost three bands are indicated.
}
\label{fig_wannier}
\end{figure*}

As a representative example, we show in Fig.~\ref{fig_wannier}a the Wannier orbitals obtained by this method for $\theta = 3^\circ$ in the $\tau=+$ spin-valley sector using the continuum model of Ref.~\cite{wang2023fractional}; those in the $\tau=-$ sector follow from time reversal conjugation. Two of them, first described in Ref.~\cite{devakul2021magic}, are almost fully layer polarized, with their centers forming the XM-MX honeycomb lattice. 
The third orbital, also found in Refs.~\cite{qiu2023interaction,xu2023maximally}, is located on the triangular MM lattice site at the centers of the previous hexagons. It has equal population between the two layers and should be understood as the interlayer bonding configuration due to the strong interlayer hopping at the MM stacking region.

\subsection{Explicit set of parameters}

Least square fits of all coefficients in our tight binding model Eq.~\ref{eq_fullTB} onto the band structure of the continuum model from Ref.~\cite{wang2023fractional} leads to the results presented in Fig.~\ref{fig_wannier}b, which we will use in the numerical results presented in this paper. We use the parameters determined from the Wannierization procedure described above as initial guess for the fit. We find that the two dominant tunneling coefficients have range $a_m$ and correspond to nearest-neighbor honeycomb-honeycomb ($t_{\rm HH}^{(1)}$) and triangular-honeycomb (($t_{\rm TH}^{(1)}$) tunnelings. At low angles, both of these coefficient are found to scale as $\theta^2$, as stressed by the quadratic fit shown with dotted lines in Fig.~\ref{fig_wannier}b. This is, in fact, very natural: the tunneling coefficient scale with the typical moir\'e kinetic energy scale $E_{\rm kin} = (\hbar \theta)^2 / 2 m^* a_0^2$, with $m^*$ and $a_0$ the effective mass and lattice constant of the monolayers. The scale of $\delta$ agrees with previous reports of the electrostatic energy difference between MM and XM-MX stacking regions~\cite{carr2018relaxation}. In Fig.~\ref{fig_wannier}c, we have superimposed the band structures of our fitted tight binding model and of the continuum model, showing perfect agreement for the topmost two bands in the considered range of twist angles; the Chern number sequences are also correctly captured.

We have also projected the full Coulomb interaction in the local orbital basis shown in Fig.~\ref{fig_wannier}a. As detailed in App.~\ref{app_interactions}, the two leading contributions are on-site interactions $U_{\rm T/H}$, and the nearest neighbor interactions connecting triangular and honeycomb sites. The latter are composed of a density-density repulsion $V$, and a chiral exchange term $J$ whose phases winds twice as fast as the single particle $t_{\rm TH}^{(1)}$ terms. As for the tight binding model, this peculiar phase structure is just a consequence of the non-zero angular momentum of the triangular orbitals (see discussion below Eq.~\ref{eq_c3eigvals}). Denoting as $\uparrow/\downarrow$ the spin component locked to the valley index $\tau = +/-$, we can write these interaction terms as 
\begin{align}
\mathcal{H}_{\rm int} & = U_{\rm H} \sum_{r \in {\rm H}} n_{r,\uparrow} n_{r,\downarrow} + U_{\rm T} \sum_{r \in {\rm T}} n_{r,\uparrow} n_{r,\downarrow}  \\
& + \sum_{\langle r, r' \rangle_{\rm TH}}  V n_{r} n_{r'} + J \left[ e^{i\phi_{r,r'}} c_{r,\uparrow}^\dagger c_{r,\downarrow} c_{r',\downarrow}^\dagger c_{r',\uparrow} + hc \right] , \notag 
\end{align}
where $\langle r, r' \rangle_{\rm TH}$ labels nearest neighbor pairs for which $r$ and $r'$ respectively belong to the triangular (T) and honeycomb (H) lattice, and the phase $\phi_{r,r'}$ grows by $2\pi/3$ starting from $2\pi/3$ as it goes through the H neighbors of a T site clockwise beginning from the rightmost one.

Only the on-site interaction were found larger or comparable to the typical kinetic energy scale $E_{\rm kin} = (\hbar \theta)^2 / 2 m^* a_0^2$ of the moi\'e structure, and we only consider those in the rest of this paper to illustrate the correlation physics emerging in the bilayer. The honeycomb orbitals being more localized than the triangular ones (Fig.~\ref{fig_wannier}a), we expect $U_H > U_T$ and have indeed found $U_H \simeq 3 U_T$ in the range of twist angles considered (see App.~\ref{app_interactions}). 
Another consequence of the more localized nature of the honeycomb orbitals is that nearest-neighbor interactions between triangular and honeycomb sites are generically larger than over those between neighboring honeycomb sites. 
This contrasts with earlier two-orbital models used in other works to describe the interacting phases of tTMDh, in which a large nearest-neighbor density-density interaction was assumed between honeycomb sites~\cite{crepel2021new,crepel2022unconventional,crepel2023topological,crepel2024attractive}. We show that this assumption misses the fine details of the correlated physics actually taking place in the material.

\section{Limiting behaviors: Chern sequences,topological transition and magic angle} \label{sec_limitingcases}

In this section we provide simple arguments for the topological transition from the spin Chern sequence $(-1,1,0)$ to $(-1,-1,2)$ as the twist angle increases, and for the existence of a magic angle at which the topmost band becomes extremely flat. These two phenomena are shown to be generic consequences of the model. 

Let us highlight two crucial features of our model that can be seen in Fig.~\ref{fig_wannier}: $(A)$ the most important tunneling coefficients have range $a_m$ and are proportional to $\theta^2$ as stressed by quadratic fits shown with dotted lines, \textit{i.e.} they scale with the typical moir\'e kinetic energy scale $E_{\rm kin} = (\hbar \theta)^2 / 2 m^* a_0^2$, with $m^*$ and $a_0$ the effective mass and lattice constant of the monolayers; and $(B)$ the triangular to honeycomb potential energy difference $\delta \sim 20-\SI{30}{\milli\electronvolt}$ is only weakly dependent on the twist angle, its value is rather fixed by the overall moir\'e depth. The very mild assumptions $(A)$ and $(B)$ together with the general form of the tight binding model Fig.~\ref{fig_bilayer}b derived from the symmetry properties of the bilayer alone confer a universal character to our arguments, which allows us to draw the generic phase diagram of tTMDh given in Fig.~\ref{fig_sketchresults}.

\subsection{Haldane limit at small angles}

\begin{figure}
\includegraphics[width=\columnwidth]{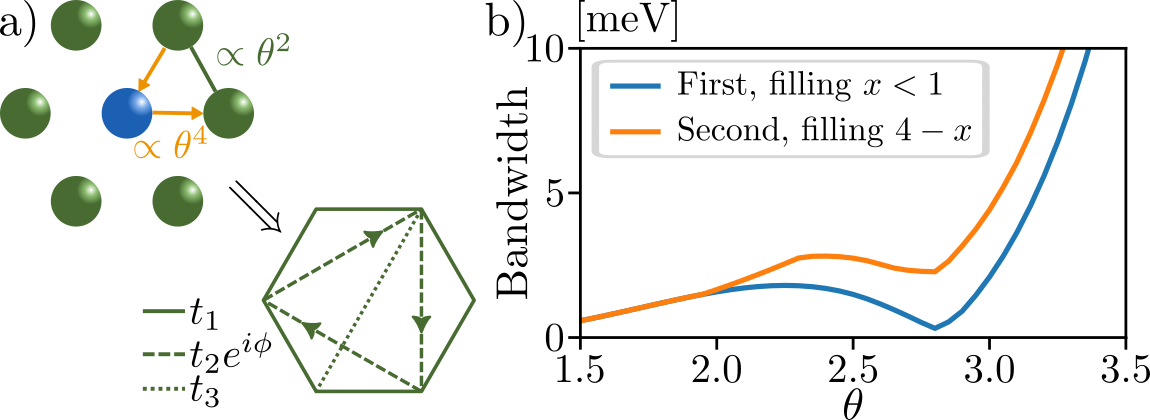}
\caption{a) The adiabatic elimination of the triangular site, with large potential energy $(-\delta)$, yields a $t_{1,2,3}$ honeycomb lattice model with complex second neighbor hopping of the Kane-Mele type with phase $\phi = 2\pi/3$, sketched here for the $\tau =+$ valley. b) Bandwidth and magic angle of the topmost valence band in the full tight binding model (blue), and of the second topmost valence band after adiabatic elimination of empty triangular sites (orange), respectively relevant for $x<1$ and $4-x$ hole doping.}
\label{fig_OrbInt}
\end{figure}

For simplicity and illustration, we truncate the tight binding model to only include the dominant (nearest neighbor) tunneling terms ($h_{\rm lr} =0$). 
Assumptions $(A)$ and $(B)$ above imply that all tunneling terms in $\mathcal{H}_{\rm TB}$ become smaller than $\delta$ for sufficiently small twist angles. 
In this regime, we can adiabatically eliminate the triangular site to reach an effective model on the honeycomb lattice describing the low-energy physics of the bilayer. This procedure, sketched in Fig.~\ref{fig_OrbInt}a, leads to an effective Hamiltonian composed of first, second and third nearest neighbor hoppings $t_{1,2,3}$, with $t_{1,3}$ real and $t_2$ of the Kane-Mele type~\cite{kane2005z} with phase $\phi = 2\pi/3$ inherited from those of $t_{\rm TH}^{(1)}$. The amplitude of $t_{1,2,3}$ in the previous model follow from standard second-order perturbation theory and read $t_1 = t_{\rm HH}^{(1)} + [t_{\rm TH}^{(1)}]^2/\delta$, $|t_2|= - t_3 = [t_{\rm TH}^{(1)}]^2/\delta$.

In absence of $t_2$, this model features two Dirac cones at the $\kappa$ and $\kappa'$ corners of the mBz. As in the Haldane model~\cite{haldane1988model}, the addition of $t_2$ with phase $\phi = 2\pi/3$ opens topological gaps at these points leading to Chern numbers $(-1,1)$ for the two topmost band. The eliminated triangular band cannot possess any Chern number as the total Chern number of our tight binding model must sum to zero. This justifies the spin-Chern number sequence $(-1,1,0)$ observed at small twist angles in Fig.\ref{fig_wannier}c.

\subsection{Magic angle in intermediate regime}

The honeycomb $t_{1,2,3}$ model with $\phi = 2\pi/3$ (Fig.~\ref{fig_OrbInt}a) possesses an isolated band with spin-valley Chern number one with minimal bandwidth and flatness ratio $\simeq 0.22$ when $t_1 = - 4 t_3$. Because of $(A)$, this flat band condition is necessarily satisfied as the twist angle of the bilayer increases from zero. For very small angles, second order corrections to $t_{1,2,3}$, proportional to $\theta^4$, are negligible and $|t_1| \gg |t_{2,3}|$. The system behaves as a honeycomb nearest neighbor tight binding model with a weak topological gap (see Fig.~\ref{fig_wannier}a at $2^\circ$). As the twist angle increases, the second order processes grow more rapidly than the first order tunneling $t_{\rm HH}^{(1)}$. As a result, there necessarily is a point where the previous magic angle condition is met (see Fig.~\ref{fig_wannier}c at $3^\circ$). Finally, for even larger twist angles, $E_{\rm kin}$ becomes comparable to the moir\'e depth $\delta$ and the system features relatively dispersive bands admixing all three orbitals (see Fig.~\ref{fig_wannier}c at $5.5^\circ$). The interference between first-order $\propto \theta^2$ and second-order $\propto \theta^4$ is clearly visible in the non-monotonic behavior of the bandwidth of the fitted tight binding model including all the terms depicted in Fig.~\ref{fig_wannier}b, as shown in Fig.~\ref{fig_OrbInt}b.

The above argument predicts the emergence of a flat band as a result of path interference through MM sites, and holds as long as the doping $x<1$ is small enough to avoid band reconstruction. It can be understood as the microscopic realization of the field-theory justification for band flattening given in Ref.~\cite{crepel2023chiral}. A generalization of this result can be obtained at filling $4-x$, \textit{i.e.} near full-filling of the topmost two moir\'e bands. After particle-hole conjugation, this regime can be mapped onto our original tight binding at filling $x$ up to an overall sign flip of the Hamiltonian. Second-order processes also change sign since second order perturbation should now be applied with an empty triangular sites. All signs being flipped, the model exhibits the same interference effects and will also feature a magic angle where the second topmost valence band flattens. This is shown in Fig.~\ref{fig_OrbInt}b, where we have explicitly integrated out the triangular site and computed the bandwidth of the second topmost valence band as a function of twist angle. This suggests that interesting correlated physics, similar to that observed below unit filling, could arise for doping levels close to four holes per moir\'e unit cells, thereby opening a new possibilities for the search of correlated phases. 

\subsection{Hofstadter limit at large angles}

From $(A)$, we know that the bandwidth generated by the tunneling terms $h_{\rm nn}$ increases as $\theta^2$ and eventually becomes comparable to or larger than $\delta$. To understand the large bandwidth limit, let us simply set $\delta = 0$ and study the topological properties of our tight binding model. In this limit, our model reduces to a tight binding model on a triangular lattice of smaller lattice constant comprising both honeycomb and triangular sites. The phases depicted in Fig.~\ref{fig_bilayer}b show that each elementary triangle of this smaller lattice is threaded by $\tau \pi/3$ units of flux. In the large angle limit, the system therefore realizes a time-reversal symmetric $\pi/3$-Hofstadter model on the triangular lattice. This model is known to possess the Chern sequence $(-1,-1,2)$; justifying the large0angle results from Fig.~\ref{fig_wannier}c.

\subsection{Summary}

We summarize the understanding obtained in this section of the topological transition occurring in tTMDh. At low angle, the spin-Chern sequence $(-1,1,0)$ appears due to the realization of an effective Kane-Mele model, while larger angle must have $(-1,-1,2)$ due to a close resemblance with the $\pi/3$-Hofstadter model. The transition between the two requires a band inversion between second and third band and occurs when the bandwidth generated by the nearest neighbor tunneling terms $h_{\rm nn}$ in Eq.~\ref{eq_fullTB} becomes comparable with the moir\'e depth $\delta$. The arguments of this section are combined to produce the general phase diagram of tTMDh presented in Fig.~\ref{fig_sketchresults}. 

\section{Short-ranged correlated phases for one hole per unit cell \label{sec:correlatedphases}} 

In this section, we focus on on-site interactions at integer filling, which we treat in a Hartree-Fock approximation for simplicity and show that our three band model reconciles, within a unified framework, the experimental observations in tTMDh at small and large angles. We also examine magneto-electric couplings in the antiferromagnetic phases of our model. 

\subsection{Hartree-Fock phase diagram}

\begin{figure}
\includegraphics[width=\columnwidth]{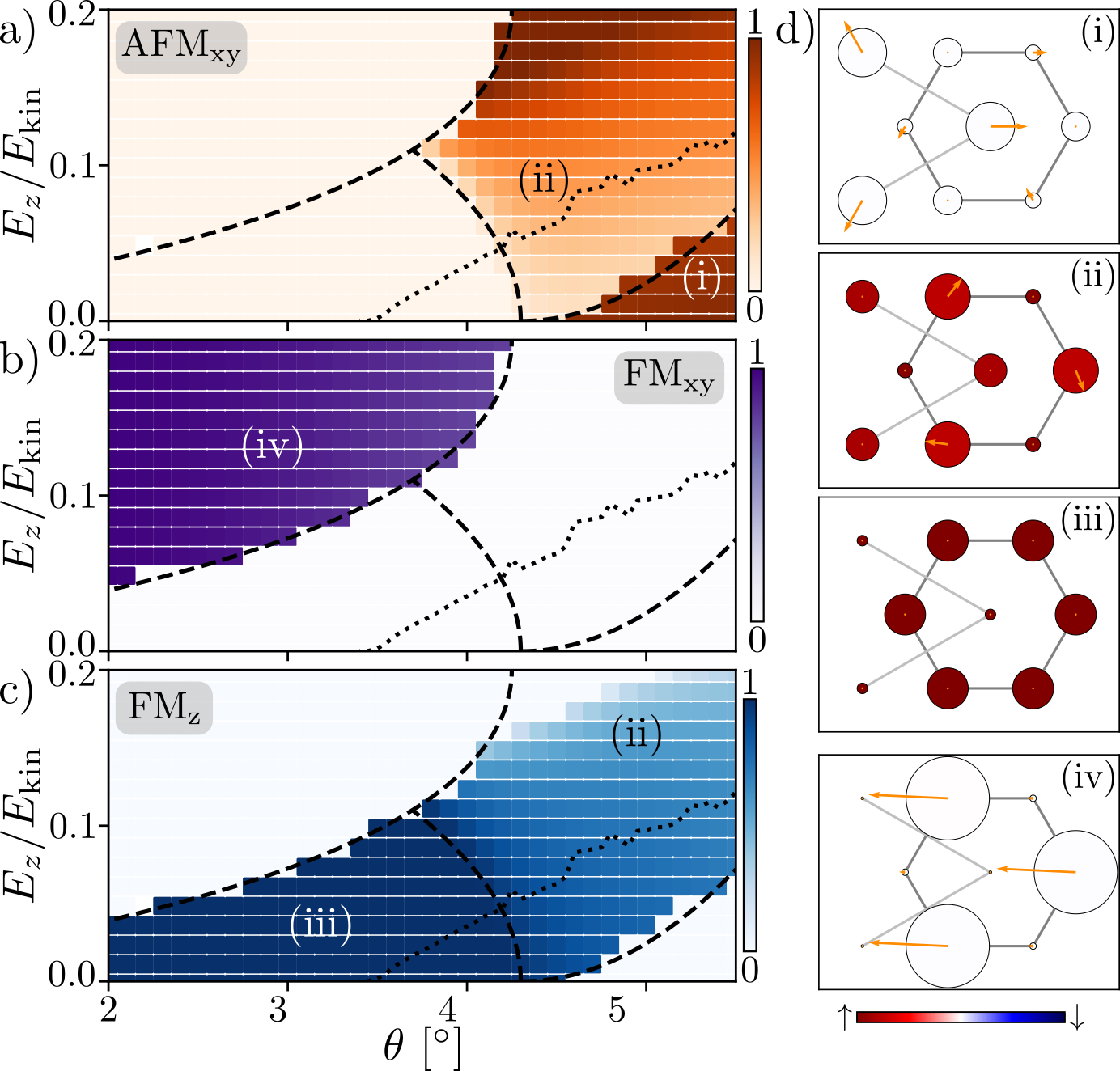}
\caption{
Panels a)-c): three representations of the phase diagram describing Hartree-Fock ground states for our three-orbital model at filling $n=1$ as a function of twist angle and displacement field, obtained using $U_{\rm T/H} = 3 E_{\rm kin}$ and $V=J=0$. 
Phase boundaries are indicated by dashed lines. The phases are characterized by three order parameters: in-plane $120^\circ$ antiferromagnetic [${\rm AFM}_{\rm xy} = \sum_{r} |S_r + \omega S_{r+a_1} + \omega^2 S_{r+a_2}|$, magnitude shown in panel (a)], in-plane ferromagnetic [${\rm FM}_{\rm xy} = \sum_{r} |S_r + S_{r+a_1} + S_{r+a_2}|$, magnitude shown in panel (b)] and out of plane ferromagnetic [${\rm FM}_{\rm z} = \sum_{r} S_r^z$, magnitude shown in panel (c)]. The dotted line gives the $E_z$ for which the van Hove singularity of the topmost valence band crosses the Fermi energy. Panel (d) shows the charge density (size of circles) and spin orientation (arrows) for order parameters shown in the different regions of panels (a)-(c). 
}
\label{fig_HF}
\end{figure}

We have determined the ground state of our three orbital model in presence of $U_{\rm T/H}$ within the Hartree-Fock (HF) approximation at the filling of one hole per moir\'e unit cell ($n=1$) as a function of twist angle $\theta$ and displacement field $E_z$, which enters our model as a potential difference between the two layers. Our results, gathered in Fig.~\ref{fig_HF}, show distinctive behaviors in the three regimes identified above (weak Kane-Mele, flat band and full orbital admixture -- see Fig.~\ref{fig_wannier}c). 

In the flat band regime around $\theta = 3^\circ$, the ferromagnetic solution with complete out-of-plane polarization has no interaction energy, only minimal kinetic energy due to the extremely small bandwidth of the topmost valence band, and therefore minimizes the full HF energy. When this ferromagnet completely fills a band with non-zero spin-Chern number, it realizes a Chern insulator and exhibits a quantized anomalous Hall (QAH) effect. This ferromagnetism and its anomalous Hall behaviors have been predicted to extend to $n<1$ and form a half-metal, providing a precursor to FCI phases~\cite{crepel2023anomalous}. All of these features were experimentally observed around $\theta \sim 3.7^\circ$ in bilayer MoTe$_2$~\cite{cai2023signatures,zeng2023integer}. 

Our calculations find a similar ferromagnetic ground state for small out-of-plane displacement fields $E_z$, which imposes a sublattice potential difference between the two inequivalent MX-XM sites with opposite layer-polarization (see Eq.~\ref{eq_fullTB}). This acts as a trivial mass term that quickly overcomes the weak Kane-Mele gap. When this mass term is large, we observe a first order transition towards an in-plane ferromagnet. All these features agree with previous works~\cite{qiu2023interaction} and can also be obtained using a two-orbital description of the bilayer~\cite{devakul2021magic}.

More interesting is the regime of higher twist angles $\theta>4.5^\circ$ where all three orbitals are strongly admixed in the topmost band. This is the regime studied experimentally in Ref.~\cite{wang2020correlated,ghiotto2021quantum}, where the nature of the correlated state at integer filling was observed to be strongly dependent on the position of a van Hove singularity in the topmost band, and could be consistently understood using a triangular lattice model. Both of these features are reproduced in our HF calculations. First, we observe that the contribution of MM orbitals to the total density of the ground state is larger than that of the honeycomb sites [phase (i) in Fig.~\ref{fig_HF}d]. We also find that the points where the HF solution has the largest gap closely follows the line where the van Hove singularity of the topmost valence band reaches the non-interacting Fermi level at integer filling. Along this line, the solution is found to be a weakly canted antiferromagnet. 

\subsection{Magneto-electric effects}

In this subsection we examine the antiferromagnetic phases and their associated magneto-electric coupling more deeply. The antiferromagnetic HF solutions stabilized at higher twist angles are characterized by their scalar $\chi^z = S_r \cdot (S_{r+a_1} \times S_{r+a_2})$ and vector $\Lambda^z = [S_{r-a_1} \times S_{r}+S_{r} \times S_{r+a_2} + S_{r+a_2} \times S_{r-a_1}]^z $ spin chiralities~\cite{grohol2005spin}, together with a potential layer polarization $\ell^z$. Here, $S_r$ denotes the magnetization of the HF ground state at site $r$.

We first consider the $E_z = 0$ limit, where the HF solution exhibits in-plane 120$^\circ$ antiferromagnetic order ($\chi^z=0$) and is two-fold degenerate to account for the two possibilities $\Lambda^z>0$ and $\Lambda^z<0$. While this HF solution spontaneously breaks time-reversal symmetry $\mathcal{T}$ due to the magnetic ordering, it preserves the $C_{2y}\mathcal{T}$ symmetry of the bilayer, \textit{i.e.} the combination of $\mathcal{T}$ with a twofold rotation around the $y$-axis $C_{2y}$. This operation exchanges the HF solutions with opposite $\Lambda^z$, and also flips the sign of $\ell^z$. As a result, the system can acquire a non-zero expectation value of $\Lambda^z \ell^z$, hence featuring a non-zero layer polarization even for $E_z=0$ due to the vector chirality of the 120$^\circ$ N\'eel state. This can be noticed in phase (i) of Fig.~\ref{fig_HF}d where the two honeycomb sites possess different particle densities (denoted as differently sized circles), which can also be observed in results from Ref.~\cite{qiu2023interaction}. Note that $\chi^z$ is preserved under $C_{2y}\mathcal{T}$ preventing its appearance in absence of applied displacement field. When $E_z \neq 0$, a similar magneto-electric effect leads to the canted antiferromagnet observed in Fig.~\ref{fig_HF} [phase (ii)]. In this case, $C_{2y}$ is broken by the applied field and the system polarizes into one layer ($\ell^z \neq 0$). We may however rely on the combination $C_{2y} \tilde{\mathcal{I}}$ with $\tilde{\mathcal{I}}$ the emergent spinless inversion symmetry found in Ref.~\cite{zang2021hartree}, which preserves $\ell^z$ and sends the HF solution into a degenerate one with opposite $\Lambda^z$ and $\chi^z$. Following the same argument, the ground state can develop a non-zero expectation value of $\Lambda^z \chi^z$, which drives the canting of the magnetic order. A similar magneto-electric coupling could be the origin of the spin canting observed in TMD heterobilayers where it leads to a transition from an in-plane antiferromagnet to a QAH insulator~\cite{li2021quantum,tao2022valley}. We aim to address this question in a future work dedicated to heterobilayers.

The various features of our HF results were previously obtained in calculations pertaining specifically to either small~\cite{devakul2021magic,qiu2023interaction} or large~\cite{qiu2023interaction,zang2022dynamical,zang2021hartree} angle tTMDh. The approach presented here provides a unified framework that very naturally captures all experimentally observed phases using a common and rather simple short-ranged tight binding model and provides a simple foundation for future theoretical work.

\section{Conclusion \label{sec:conclusion}}

Relying solely on the geometry of twisted TMD homobilayers and their spin-valley locking, in this paper we have derived a tight binding model for these heterostructures, which encompasses their physics in the full range of twist angle and moir\'e depth. 
This description gives the first microscopic explanation for the magic angles in terms of an orbital path interference, and interpolates between the honeycomb localized Kane-Mele physics of small twist angles and the triangular Hofstadter regime realized at large angles. 

The different regimes identified can be used as a guide to get closer to an experimentally interesting situation. For instance, one should increase the twist angle to reach the magic point of small bandwidth if a crossing at $\kappa$ is detected in the band structure by ARPES, or if a strongly localized honeycomb density profile is measured using STM.

The natural local basis identified here respects the topological data extracted from large-scale \textit{ab-inito} calculations and perfectly captures the band-structure of effective continuum models. It enables the investigation of interaction effects in the topological band of twisted TMDs. This is exemplified by our Hartree-Fock calculations at integer filling that correctly capture in a unified framework the various phases and magneto-electric properties observed experimentally from low to large twist angles. This local basis seems particularly well suited for the study of correlation effect for filling above one hole per moir\'e unit cell, a regime that is difficult to access using band projected method due to strong renormalization of the band structure caused by the small inter-band gaps. Note however that the third Wannierized hole band is hardly ever isolated from lower valence moir\'e bands that our model discards, indicating a possible breakdown of our model for fillings above four holes per moir\'e unit cells. 

Finally, we hope that the new chiral exchange interaction identified may help understand the robustness of the experimentally observed FCI phases using methods similar to decomposition into pseudo-potentials~\cite{haldane1983fractional,haldane1988spin,crepel2019matrix}. Extension of our work to the interaction and topological properties of hetero-bilayers will be addressed in a future study.

\section*{Acknowledgements} The Flatiron Institute is a division of the Simons Foundation. VC greatly benefits from discussions with R. Queiroz and N. Regnault on the physics of moir\'e heterostructures. VC thanks X.W. Zhang and D. Xiao for providing additional details on their \textit{ab-initio} calculations. AJM acknowledges support from the National Science Foundation (NSF) Materials Research Science and Engineering Centers (MRSEC) program through Columbia University in the Center for Precision Assembly of Superstratic and Superatomic Solids under Grant No. DMR-1420634.

\appendix

\section{Chiral exchange} \label{app_interactions}

\begin{figure}
\includegraphics[width=0.8\columnwidth]{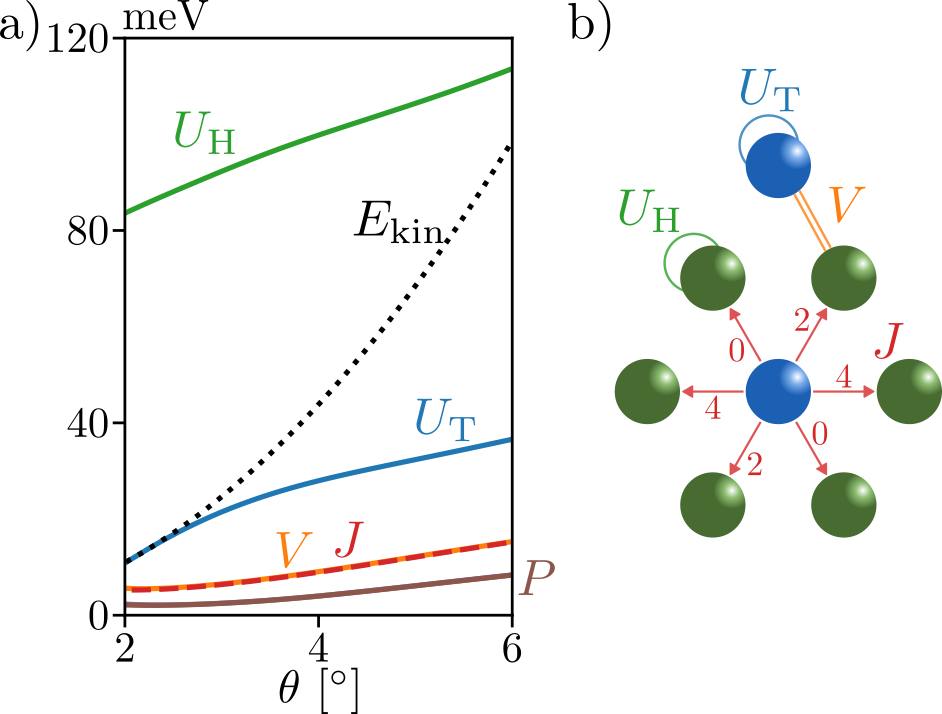}
\caption{a) Amplitude of the dominant $U_{\rm H}, U_{\rm T}$ interactions terms, subleading $V, J$ and of all other neglected one that remain smaller than the pair hopping scale $P$. b) Sketch of the interaction terms of the lattice, with the winding phases of the exchange depicted by arrows.}
\label{Fig_Interactions}
\end{figure}

To complete our description of the bilayer, we project the Coulomb interaction potential $V(r) = e^2 / (4\pi \varepsilon r)$, with $\varepsilon$ the monolayer dielectric constant, onto the orbitals obtained in Fig.~\ref{fig_wannier}a. Unsurprisingly, we find that the on-site interaction between opposite spin-valley flavors largely dominates over all other interaction terms and is larger on the more localized honeycomb orbitals; $U_{\rm H} \sim 90$meV and $U_{\rm T} \sim 30$meV. After these dominant terms, we see that only interactions between nearest triangular-honeycomb neighbors are appreciable, and follow the following hierarchy. Density-density interaction $V \sum_{\tau, \tau' = \pm} n_{r,\tau} n_{r',\tau'}$ and exchange terms $J c_{r,+}^\dagger c_{r,-} c_{r',-}^\dagger c_{r',+}$ are of similar magnitude $\sim 10$meV, and largely dominate over all the other terms such as pair hopping $P c_{r,+}^\dagger c_{r,-}^\dagger c_{r',-} c_{r',+}$ with typical scale $\sim 2$meV. This is summarized in Fig.~\ref{Fig_Interactions}a, where we plot the values of these coefficients as a function of twist angles. All other interaction terms on the lattice, like $P$, were found negligible compared to $V$ and $J$. Note that having $V$ and $J$ of similar order is rather uncommon, the electrostatic density-density interaction should indeed dominate the quantum tunneling mediated exchange for large inter-orbital separations. 
We recover this more standard behavior for nearest neighbor interactions between honeycomb sites and for all interaction with range $>a_m$, for which density-density interactions largely dominate the other terms.

The most interesting feature of these interactions is subtly hidden in the phases involved in the exchange interaction, which are represented in Fig.~\ref{Fig_Interactions}b and can be formally described as 
\begin{align} \label{appeq_interactions}
\mathcal{H}_{\rm int} & = U_{\rm H} \sum_{r \in {\rm H}} n_{r,\uparrow} n_{r,\downarrow} + U_{\rm T} \sum_{r \in {\rm T}} n_{r,\uparrow} n_{r,\downarrow}  \\
& + \sum_{\substack{r \in {\rm H} \\ r' \in {\rm T}}} \!\! V n_{r} n_{r'} + J \left[ e^{i\phi_{r-r'}} c_{r,\uparrow}^\dagger c_{r,\downarrow} c_{r',\downarrow}^\dagger c_{r',\uparrow} + hc \right] , \notag 
\end{align}
with the phase $\phi_{r-r'}$ grows by $2\pi/3$ starting from $2\pi/3$ as it goes through the MX-XM neighbors of a MM site clockwise beginning from the rightmost one (the phase are twice those of the single-particle terms in Fig.~\ref{fig_bilayer}b as there are twice as many fermionic operators). This very peculiar structure is inherited from the momentum of spin-valley locked holes. The phases in Eq.~\ref{appeq_interactions} are simply twice those of the single particles terms (see Fig.~\ref{fig_bilayer}b), as the interactions involves twice as many fermionic operators.

the orbital character difference $s/p_\pm$ of the MM/XM-MX orbitals. After spontaneous spin-valley polarization of the system, the on-site terms have no effects on the physics anymore, and the chiral winding of the exchange may have profound physics of the bilayer.

\bibliography{TBmote2}

\end{document}